\documentclass{aastex}
\usepackage{emulateapj5,apjfonts}
\newcommand{\quan}[3]{\ensuremath{\rm #1^{+#2}_{-#3}}}

\newcommand{\Nh}{\mbox{${\rm N_{H}}$}}
\newcommand{\ecs}{\mbox{$\rm erg\;cm^{-2}s^{-1}$}}
\newcommand{\es}{\mbox{$\rm erg\;s^{-1}$}}

\newcommand{\kev}{\mbox{\rm keV}}

\newcommand{\mr}{\ensuremath{\rm m_R}}
\newcommand{\mv}{\ensuremath{\rm m_V}}
\newcommand{\mb}{\ensuremath{\rm m_B}}
\newcommand{\et}{{\it et al.\/}}
\newcommand{\CXO}{{\it CXO}}
\newcommand{\ACIS}{{\it ACIS}}
\newcommand{\EGRET}{{\it EGRET}}
\newcommand{\ASCA}{{\it ASCA}}
\newcommand{\GIS}{{\it GIS}}

\newenvironment{inlinefigure}{
\vspace{0.35truecm}
\def\@captype{figure}
\noindent\begin{minipage}{0.999\linewidth}\begin{center}}
{\end{center}\end{minipage}\smallskip}

\shorttitle{Chandra Imaging of the Gamma-Ray Source GeV J1809-2327}
\shortauthors{}

\begin{document}

\title{Chandra Imaging of the Gamma-Ray Source GeV J1809-2327}

\author{Timothy M. Braje, Roger W. Romani}
\affil{Department of Physics, Stanford University, Stanford, CA 94305}
\email{timb@astro.stanford.edu, rwr@astro.stanford.edu}

\author{Mallory S.E. Roberts}
\affil{Department of Physics, McGill University, 3600 University St. 
Montreal, QC. H3A 2T8 Canada}
\email{roberts@hep.physics.mcgill.ca}

%\and

\author{Nobuyuki Kawai}
\affil{
Department of Physics, Tokyo Institute of Technology, Ookayama,
Meguro-ku, Tokyo 152-8551, Japan \\and RIKEN, 2-1 Hirosawa, Wako,
Saitama 351-0198, Japan}
\email{nkawai@tithp1.hp.phys.titech.ac.jp}

\begin{abstract}

We report on Chandra imaging observations of the Galactic
Unidentified $\gamma$-ray source GEV J1809-2327, comparing the X-ray
images with new VLA 1.46 GHz and 4.86 GHz maps. The X-ray images
reveal a point source connected to a non-thermal X-ray/radio nebula,
supporting a pulsar/wind model for the
$\gamma$-ray emitter. We also detect numerous X-ray sources from
the young stellar association in the adjacent HII region S32.
\end{abstract}


\section{Introduction}

The \EGRET\ instrument on the {\it Compton Gamma Ray Observatory}
detected many bright sources along the Galactic plane, most of which
remain unidentified. A number of these sources have high significance
in $E> $GeV photons \citep{lam97}, providing relatively good source
localizations. The key to making progress on identifications is to
search for lower energy, particularly hard X-ray and radio,
counterparts. \citet*[RRK]{rob01} have carried out an \ASCA\ survey
of the bright GeV sources to search for counterparts and test the
nature of the GeV emitters.
\bigskip

%low levels (tstsm.fits 12.4:16+3) high levels (cxoadap.fits 1:50+1)
\begin{inlinefigure}
\epsscale{0.95} \plotone{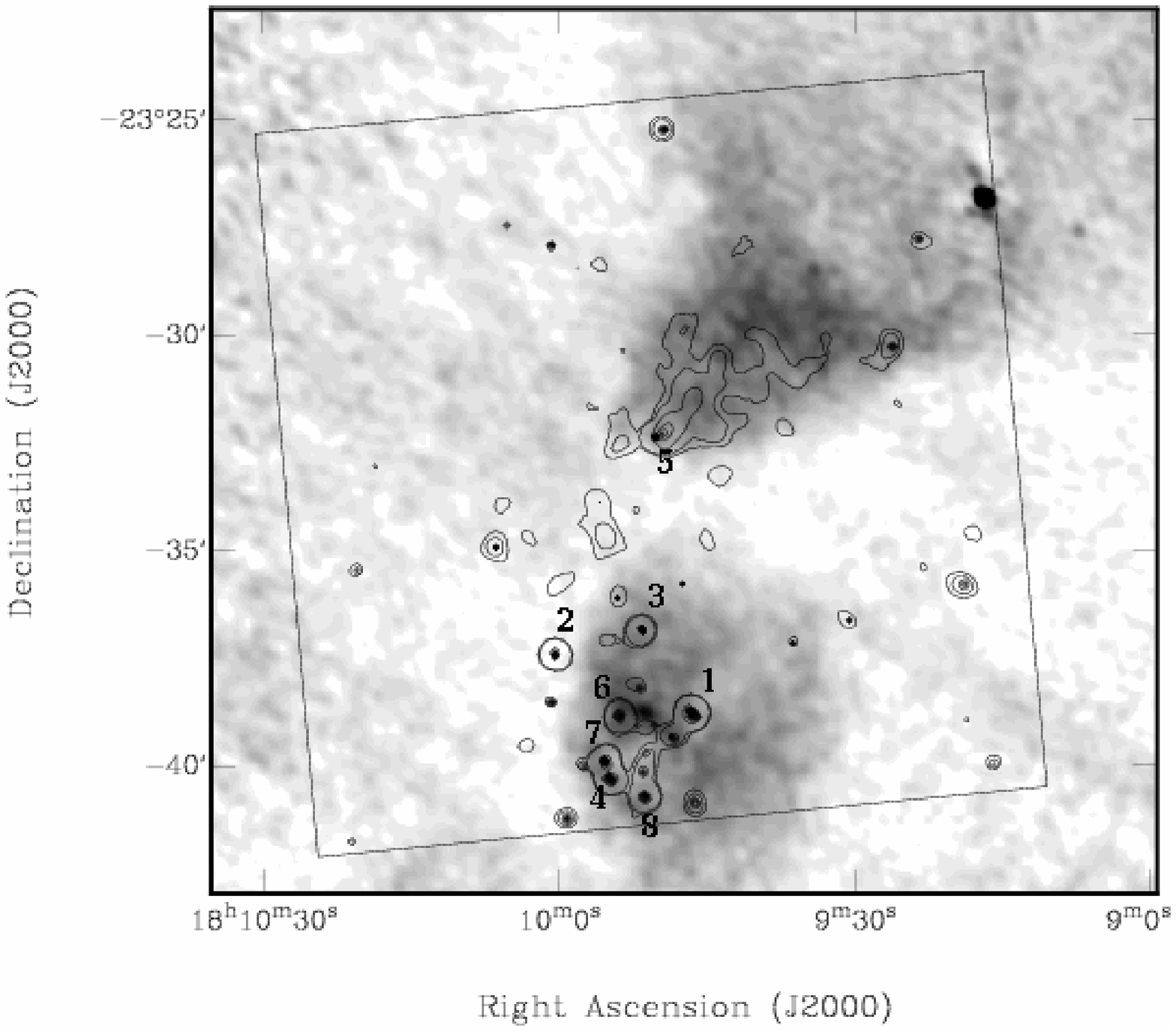}
\figcaption{VLA 21cm map with ACIS 0.5-8keV smoothed contours. The ACIS-I FOV
(square) and the brightest sources of Table 1 are also indicated.}
\label{xray-radio}
\end{inlinefigure}

GeV J1809-2327(=2EG J1811-2339) is one of the brightest,
best-localized unidentified plane sources. It is one of a subset of
sources showing evidence of $\gamma$-ray variability,
with an extended hard-spectrum X-ray counterpart argued to be a
pulsar synchrotron wind nebula, or PWN \citep{rob01}. This source
is also interesting because it is adjacent to the dark cloud Lynds 227
and to the HII region S32 with an embedded association of young
high-mass stars. These are plausibly associated with the X-ray source
and at $\sim 1.8$kpc provide a possible birth site for a pulsar.
  \citet{oka99}, in a study of the molecular gas in this
region, found morphological support for this association and
hypothesized that the $\gamma$-ray photons were produced when TeV
pulsar electrons penetrated the dense molecular gas, suffering
relativistic bremsstrahlung losses.

Studies of this complex region were hampered by the modest angular
resolution of the \ASCA\ \GIS\ ($\sim 3\arcmin$ half power diameter), 
so we have
obtained high resolution \ACIS\ images to separate out the point
source contribution to the X-ray complex. This is abetted by new VLA
continuum maps. These data support the pulsar/PWN hypothesis and
give new constraints on the origin of the high energy emission.

\section{Observations and Data Analysis}

\subsection{\CXO\ Imaging and Source Detection}

We observed the gamma-ray source GeV J1809-2327 on 25 August 2000 with
the \ACIS\ detector \citep{bur97} on the Chandra X-Ray Observatory 
\citep{wei00} for $\sim
9.7$ks. The exposure had the imager chips I0-3 at primary
focus with additional off-axis coverage from chips S2 and S3.
Integrations were made in TE mode. Basic analysis of the (reprocessed)
data was accomplished using CIAO v2.2.0.1.  We found no evidence of
background event rate flares;
therefore, the entire  dataset was used in the analysis.

Our primary X-ray source list was established using \emph{wavdetect}
with a significance threshold of $10^{-7}$.   Forty four sources are detected with
a quoted significance $>3.0$, and in Table~\ref{src_cts}  we list
those sources with a significance of greater than 8.0.  The most significant
of these are marked in Figure 1. Comparison of
our optical and X-ray frames allows us to boresight positions to $\sim
0.3\arcsec$.  Several of these sources have matches in SIMBAD and the
USNO A2 catalogues.  Many of these lie in the S32 stellar association
to the south of our GeV counterpart (Figure~\ref{xray-radio}).  An
overlay of the X-ray image with POSS II scans reveal roughly a dozen
additional matches, either with optical $m >20$ or lower (albeit still
high) X-ray significance.

\begin{table*}[t]
\begin{center}
\footnotesize
%\rotate
\tablewidth{0pt} 
\caption{\label{src_cts}X-ray sources} 

\begin{tabular}{crrrrrrl}\hline
CXOU J Source & Src & Counts & ${\rm P_{var}}$ & kT & \Nh & ${\rm Flux_{0.5-8.0}}$ & Comments \\
\Huge
& & & (\%) & (\kev) & (\mbox{$\rm 10^{22} cm^{-2}$}) &  \multicolumn{2}{l}{( $10^{-14}$ \ecs)}  \\
\hline\hline
\footnotesize
180926.1-233017 &17& $30.5 \pm  5.9$ &  86.2 & & & & \\ %sig 9.2
180946.2-234052 &12& $  44.8 \pm  7.3$ &  35.1 & & & 
	& CD--23 13997 $\mv=9.22$ $\mb=9.21$ \\ %sig 10.8
180946.6-233846 &1& $ 227.1 \pm 15.5$ &  96.5 & \quan{1.07}{0.06}{0.08} & \quan{0.6}{0.2}{0.1} & \quan{13}{1}{1} & HD166033 $\mv = 8.61$, $\mb = 8.65$ \\ %sig 48.3
180948.4-233921 &13& $  42.1 \pm  7.1$ &  36.5 & & & 
	& HD314032 $\mv=9.90$, $\mb=10.07$ \\ %sig 10.6
180949.5-232514 &9& $  45.7 \pm  7.2$ &  98.8 & & & 
	& $\mr=14.5$, $\mb=15.1$ \\ %sig 12.5
%180950.2-233223 &5& $  85.6 \pm  9.3$ &  15.7 & \quan{0.226}{0.006}{0.007} (BB) & \quan{1.02}{0.14}{0.11} & \quan{4.22}{0.73}{0.74} & \\ %sig 34.8
180951.2-234044 &8& $ 104.0 \pm 10.7$ &  48.7 & \quan{0.71}{0.13}{0.08} & \quan{1.3}{0.2}{0.2} & \quan{4.8}{0.8}{0.8} & \\ %sig 24.3
180951.5-234009 &19& $  38.2 \pm  7.1$ &  12.9 & & & & \\ %sig 8.4
180951.6-233652 &3& $  86.9 \pm  9.4$ &  71.6 & \quan{0.5}{0.1}{0.1} (BB) & \quan{0.1}{0.3}{0.1} & \quan{6}{1}{1}
	& WD 1806--23 $\mv=9.99$, $\mb=10.7$ \\ %sig 36.7
180951.8-233811 &18& $  22.3 \pm  4.9$ &  11.1 & & & & \\ %sig 8.7
180953.9-233852 &6& $ 110.9 \pm 10.8$ &  71.2 & \quan{0.9}{3}{0.3} & \quan{1.2}{0.3}{0.8} & \quan{6.3}{0.9}{0.9} & \\ %sig 32.4
180954.2-233605 &15& $  19.8 \pm  4.5$ &  11.8 & & & & \\ %sig 9.9
180954.9-234018 &4& $ 166.3 \pm 13.3$ & 100.0 & \quan{0.59}{0.09}{0.10} & \quan{1.2}{0.2}{0.1} & \quan{7.3}{0.9}{0.9} & \\ %sig 36.7
180955.4-233954 &7& $ 142.7 \pm 12.4$ &  57.8 & \quan{0.9}{2}{0.1} & \quan{1.0}{0.2}{0.9} & \quan{7.6}{0.9}{1.0} & \\ %sig 31.9
180957.5-234001 &21& $  26.6 \pm  5.6$ &   1.2 & & & & \\ %sig 8.1
180959.2-234115 &10& $  47.8 \pm  7.4$ &  97.8 & & & & \\ %sig 12.5
181000.4-233726 &2& $ 136.3 \pm 11.8$ &  98.9 & \quan{1.0}{0.4}{0.1} & 1.1 (fixed) & \quan{7}{1}{1} & \\ %sig 43.8
181000.8-232756 &14& $  21.4 \pm  4.7$ &  53.0 & & & & \\ %sig 10.0
181000.8-233832 &16& $  23.0 \pm  4.9$ &  55.9 & & & & \\ %sig 9.9
181006.4-233456 &11& $  25.6 \pm  5.1$ &  48.5 & & & & $\mr=15.5$, $\mb=17.4$ \\ %sig 12.4
181052.5-233713 &20& $  69.7 \pm 11.1$ &  96.5 & & & & \\ \hline \hline %sig 8.3
nebula        & $\cdots$ &    &    & \quan{\Gamma=2.2}{0.4}{0.4} & \quan{2.2}{0.6}{0.5} & \quan{59}{4}{4} & \\
180950.2-233223 &5& $  85.6 \pm  9.3$ &  15.7 & \quan{0.30}{0.1}{0.09} (BB) & \quan{0.7}{0.7}{0.4} & \quan{5.0}{0.8}{0.8} & \\ %sig 34.8
                &5&                   &       & $\Gamma=$\quan{5}{2}{1} & \quan{1.3}{0.9}{0.5} & \quan{6}{1}{1} & \\ %sig 34.8
                &5&                   &       & $kT=$\quan{0.18}{0.08}{0.06} (BB+PL) & \quan{1.4}{1.4}{0.7} & \multicolumn{2}{l}{ \quan{5}{3}{3} (PL), \quan{3.4}{0.6}{0.6} (BB)} \\
%\quan{8.3}{1.5}{1.5} & \\ %sig 34.8
\hline\hline
\end{tabular}
\end{center}
\end{table*}

We tested these sources for variability, comparing the individual
source photon arrival times to the total image counts (mostly
background)  arrival times, using the  K-S statistic \citep{pre92}.
Sources with a variability probability greater than $\sim 95\%$ we
regard as likely variables, while most sources are consistent with a
constant flux.

As in earlier \ASCA\ images, it is evident that in the northern half
of the field there is also diffuse X-ray emission.  After exposure
correction and application of adaptive smoothing filters we find a
wedge of diffuse emission in our field with point source 5 at its
apex. Figure~\ref{xray-radio} shows contours from the smoothed image
overlaid on a 1.4GHz continuum map. At very low surface brightness below the
faintest contours shown, the diffuse X-ray emission seems to extend to
the lower left, which might place point source 5 towards the center,
rather than apex of the diffuse flux. However, the brightest emission
to the  NW correlates well with the radio continuum flux.
\medskip
\medskip
% can get photons in cxo_qsub.fits
\begin{inlinefigure}
\epsscale{0.9} \plotone{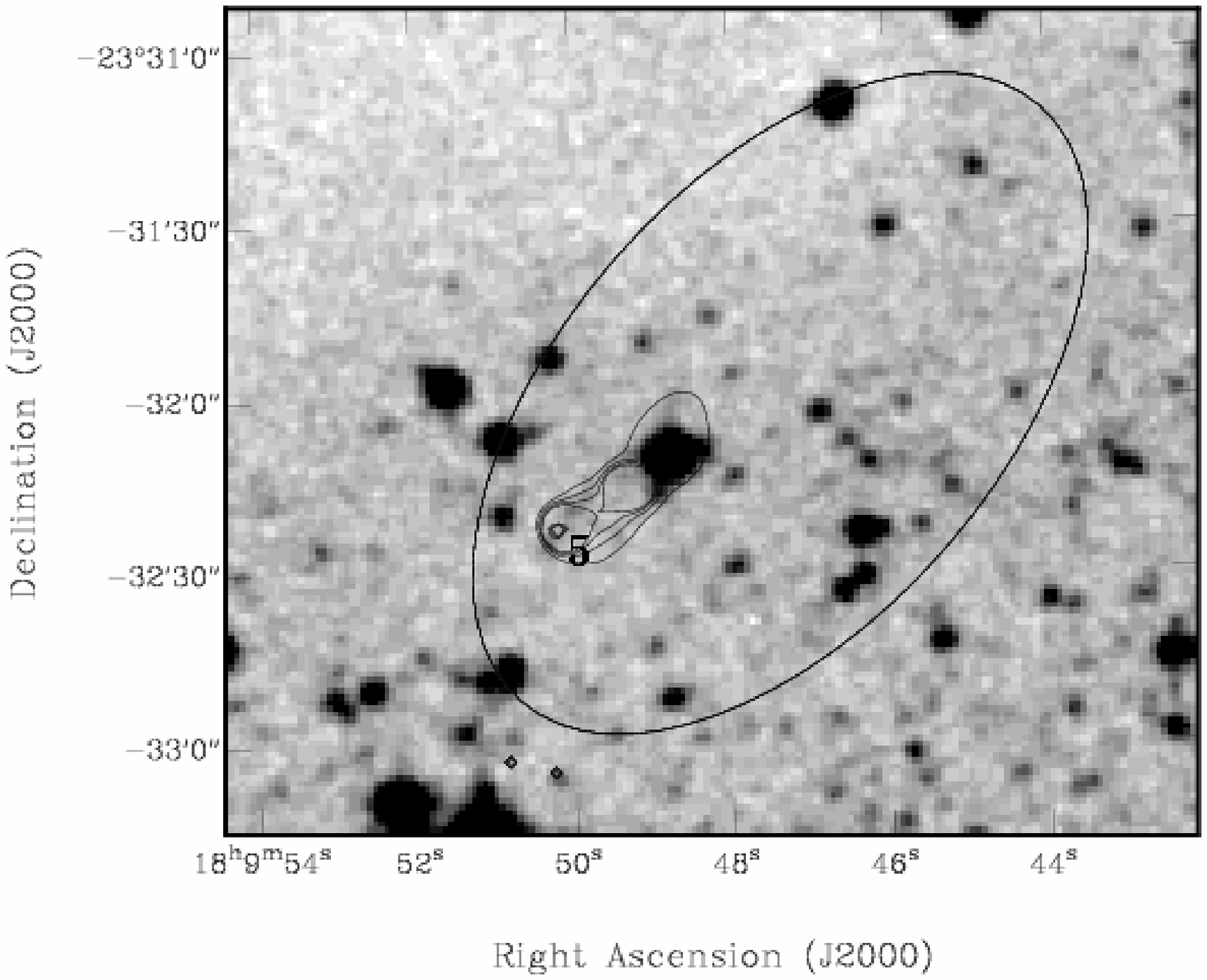}
\figcaption{DSS image around source 5. Smoothed ACIS contours show the 
point source and $30\arcsec$ `jet'. The ellipse bounds our nebular extraction region.}
\label{xray-opt}
\end{inlinefigure}

Figure 2 overlays X-ray
contours on the POSS2 R image. There is no optical point source to the
plate limit within $\sim 1.5\arcsec$ of the X-ray position.  Optical
extinction is modest at source 5, but increases rapidly  $\sim
30\arcsec$ to the NE, toward the edge of Lynds 227.
There is significant emission in a narrow jet or trail from
point source 5 to the diffuse nebula. In the raw photon counts it
shows as a narrow linear feature with a brighter knot $\sim
15\arcsec$ from the point source.

\subsection{X-ray Spectral Fitting}

For the brightest sources ($\ga~80$ counts), we have attempted
individual spectral analyses, creating PI spectra, Ancillary Response
Files, and Response Matrices.  Because of the limited counts and the
limited \ACIS\ spectral resolution, several spectral models were, in
general, formally acceptable. We chose to fit the bright stellar
sources in S32 with an absorbed Raymond-Smith plasma model.  
Fit parameters and multi-dimensional error estimates are listed in
Table~\ref{src_cts}.
\bigskip

\subsubsection{Nebula Analysis}

	Our earlier \ASCA\ observations had longer exposure (80ks) and larger 
effective area. We thus compare present spectral estimates with 
the higher statistical significance (but spatially unresolved) estimates in
RRK.  The \ASCA\ extraction aperture was 
quite large (4$\arcmin$ radius), so background (particles and
Galactic X-rays) dominates the \ACIS\ counts in this region. Accordingly,
we also define a smaller extraction region (a $68\arcsec \times 38\arcsec$
ellipse plus exclusion of the point source) covering the brightest 
part of the nebula (Fig. 2). In the larger aperture, we attempt to account for 
the particle-induced background by subtracting the 
\ACIS\ `blank-sky' images. For the smaller aperture, this blank sky
subtraction made no significant difference in the source parameters; for 
this and point source extraction regions, we skip this step.
In all cases we carefully chose background regions on the ACIS-I
chips that were free of obvious emission and spanned a similar range
of CCD rows to the source aperture.

\begin{inlinefigure}
\epsscale{0.8} \rotatebox{270}{\plotone{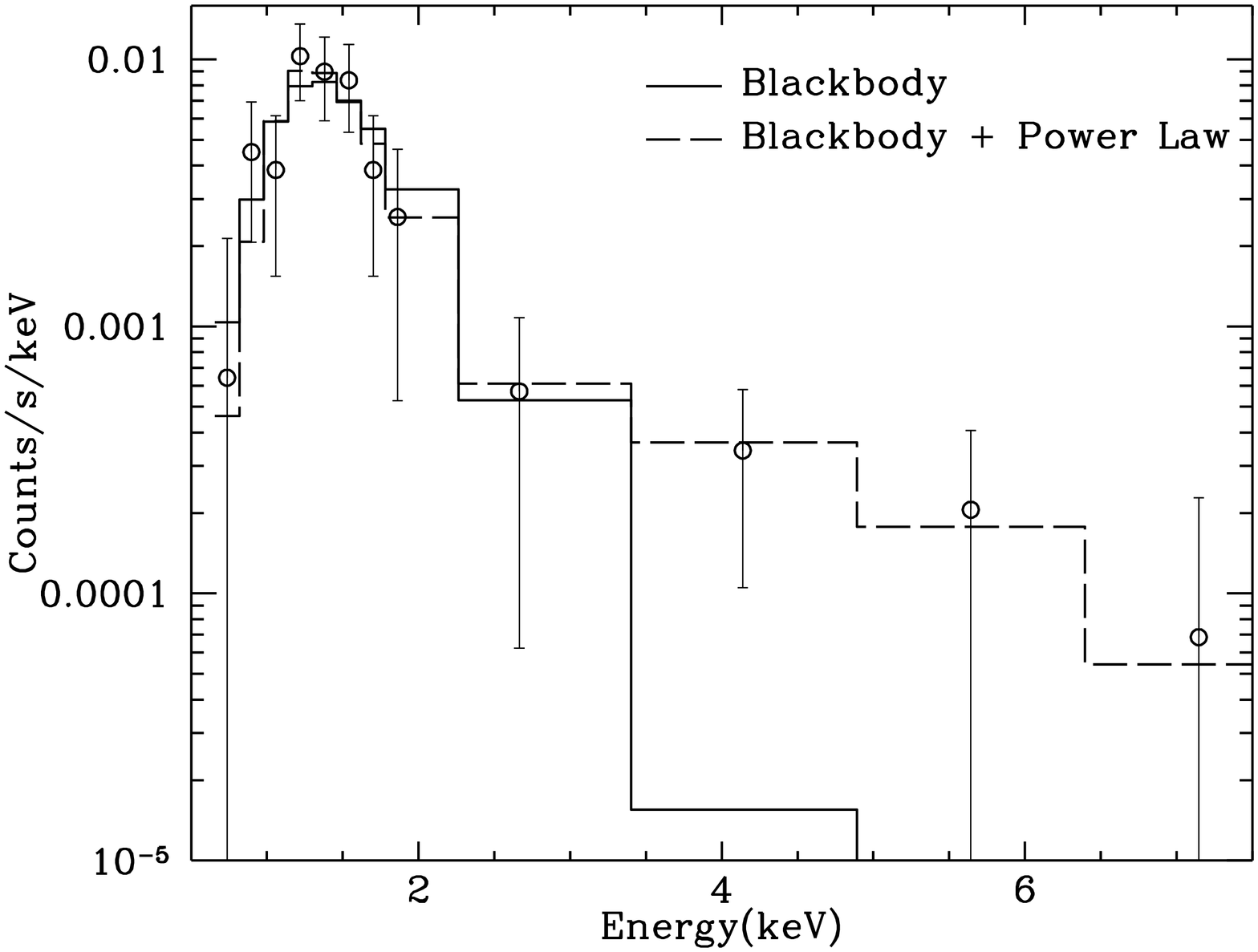}}
\figcaption{Count spectrum and models for point source 5. The excess counts
above 3 keV are highly significant and require a hard model component.}
\label{pt-spect}
\end{inlinefigure}

	\ASCA\ measurements of the diffuse emission determined
a power-law index of $\Gamma = \quan{2.09}{0.18}{0.17}$ and an absorption of
$\Nh=\quan{1.77}{0.29}{0.26} \times 10^{22} {\rm cm^{-2}}$ (90\% errors). 
Our fits to the \ACIS\ counts in the small extraction region produces values
consistent with these (Table~\ref{src_cts} -- quoted errors
are multi-dimensional from the sherpa command \emph{projection}). 
In the larger region we obtain
power-law fit parameters several sigma away from the \ASCA\ values. We
believe that, even with the blank sky subtraction, the dominant particle
contribution is not adequately removed in this large aperture. Unlike
the small aperture, we also find that the
\ACIS\ fit parameters are significantly affected by the
region chosen for the local background subtraction. If, however, we
fix the spectral parameters at the \ASCA-determined values and fit the counts in
the large aperture, we obtain $3.6 \times 10^{-12} \ecs$ for the \ACIS\ data,
in good agreement with the flux measured in \citep{rob01}. We therefore
consider that the \CXO\ data are consistent with the \ASCA\ fits,
but adopt the latter spectral indices as fiducial for the diffuse emission.

\subsubsection{Point Source Analysis}

For source 5, our counterpart point source, we fit a simple absorbed
blackbody.  While statistically acceptable, this model cannot produce
the counts above 3\kev\ (Figure~3), which given
our small $\sim 2\arcsec$ extraction aperture are highly significant
($< 0.01$ background counts and $< 1$ pile-up count are expected). 
Interestingly, several X-ray emitting young pulsars have hard power
law components (presumably magnetospheric) superposed on soft thermal spectra
\citep[e.g.][]{sai97,hal97,pav01}, with a typical index
$\Gamma=1.5$. We therefore assume this $\Gamma$ 
and fit a powerlaw amplitude from the $E>3\kev$ 
source counts. We then fix this power law and re-fit the soft BB component.
Parameters for the combined model are listed in Table 1.
%; the  power-law component flux is poorly  constrained at 
%$2\pm 2\times 10^{-14}\ecs$. All three models produce Gehrels-$\chi^2 < 0.7$ 
(with the combined fit smallest by $\sim 2 \times$),
a result of the low count statistics in the high energy bins.

%%%  Fix the significant figures in this region..... ???
%to find for the combined BB+PL model $\Nh=\quan{1.48}{0.15}{0.12}$, 
%$kT=0.170\pm 0.004$keV, and 
%${\rm flux}=7.47\pm 0.66\times 10^{-14}\ecs$ (0.5-8 keV total).  
%The power-law component contributes ${\rm flux}=3.5\pm 3\times 10^{-14}\ecs$.

\subsection{Radio Continuum Imaging}

Mosaiced continuum observations of the GeV J1809$-$2327 field at 1.46 GHz
and 4.86 GHz were performed with the VLA.
%in the D, DnC, and CnB arrays.     
%Full Stokes parameters were retained allowing linear polarization
%mapping. Spectral measurements were performed by matching the
%$uv$ coverage between the two observing bands and then producing
%spectral tomography maps. 
Details of the radio observations and full results will be presented
elsewhere (Roberts, Gaensler, and Romani in preparation).  
The 1.46 GHz field (Fig.~\ref{xray-radio}, $18\arcsec \times 15\arcsec$
beam) shows diffuse emission associated with S32 and the X-ray PWN;
both are also well covered in the 4.86 GHz mosaic.
%three nebulae in a north-south line, with
%the central and northern ones well covered by the 4.86 GHz mosaic.
%The central nebula is coincident with the S32 stellar association, and the
%northern nebula is coincident with the diffuse X-ray trail.

%From the ratios of the continuum fluxes it is clear that emission in
The S32 region has an inverted, partly thermal spectrum, with spectral index $\alpha
\approx 0.1$.  The morphology is hollow centered, following the
contours of the infrared HII region. In  contrast, the wedge of
continuum to the north has a non-thermal spectrum, with an approximate
spectral index of $\alpha_R \approx -0.3\pm0.1$ and
significant polarization.
% at 6cm (Roberts, Gaensler, and Romani 2001).
The emission follows our diffuse X-ray trail extending $\sim 6\arcmin$
to the NW, where the X-ray emission fades below detectability.

\section{Interpretation and Conclusions}

Comparing our new observations with the pulsar+PWN hypothesis, we note that
source 5 composite spectrum is characteristic of a young 
pulsar, with a hard (presumably magnetospheric) power law and 
underlying thermal emission. Given the high $kT=0.18$keV, our fit thermal flux
implies a small effective area of
$
{\rm A_{eff}}\approx 4 \times 10^{11} (d/1.8{\rm kpc})^2 {\rm cm^2}.
$
At $\la 3$\% of a neutron star's surface, this could represent a polar cap 
heated by precipitation of magnetospheric particles. Alternatively,
atmosphere effects \citep{ro87,pav01} can produce a hard thermal tail,
although such a large fit $kT$ is difficult to produce for reasonable pulsar 
ages. At a more plausible  $kT_{eff} \la 0.05\kev$, full surface emission from
cooling would be unobservable; our data only limit such a component to
$< 1.2 \times 10^6$K.

	Both the power law emission and the hot thermal component can thus
measure magnetospheric activity, and have been phenomenologically
related to the spin-down luminosity. \citet{bt97} find that 
%the ROSAT flux
for pulsars observable by ROSAT, the flux 
scaled as  $L_x(0.1-2.4\kev) \approx 10^{-3} {\dot E}$; after correcting for
absorption our inferred $0.1-2.4\kev$ flux of $\approx 2 \times 10^{-12} \ecs$
corresponds to a spin-down luminosity ${\dot E} = 8 \times 10^{35} 
d_{1.8}^2\es$ at the fiducial source distance. \citet{sai97} 
find that the (pulsed) \ASCA\ luminosity scales as $L(2-10\kev) \approx 10^{34}
{\dot E_{38}}^{3/2} \es$. Using our 2-10keV unabsorbed point source flux,
we infer from this relationship ${\dot E} = 1.5 \times 10^{36} 
d_{1.8}^{4/3} \es$.  For dipole spindown we have ${\dot E} 
\approx 10^{38} (B_{12} \tau_4)^{-2} \es$; these luminosities indicate 
typical spin parameters for a $\gamma$-ray emitting pulsar of $B_{12} 
\tau_4 \approx 10$, where the surface dipole
field is $10^{12}B_{12}$G and the pulsar age is $10^4 \tau_4$y.

The broadband spectral energy distribution (SED, Figure 4) of our PWN candidate
provides some useful constraints on its nature,
if we interpret the radio/X-ray spectrum as synchrotron emission from a
cooling electron population: ${\rm d} N_e/{\rm d}\Gamma_e = 
A \Gamma_e^{-s}$ with $\Delta s= 1$ at a break $\Gamma_B$.
This corresponding to a photon break frequency $\nu_B = 4.2 \times 10^{13}
B_{10} \Gamma_6^2$, with a typical PWN field of $10B_{10} \mu$G and
a electron break at a Lorentz factor $\Gamma_B = 10^6\Gamma_6$. 
Our SED allows a modest range
for the power law index $s=2.85\pm 0.15$. The data require a very 
tight correlation between $s$ and the break frequency,
${\rm Log} (\nu_B) = 9.1\, s -13.3$; however this allows a large 
range of break energies from the microwave band through the near IR.

	The conventional PWN picture identifies this break with the
cooling of the electrons at $\tau_4 = 19 B_{10}^{-2} \Gamma_6^{-1}$,
associating this with the pulsar characteristic age.
In this approximation (a homogenous, uniform field PWN) our
data then can be used to eliminate $\Gamma_B$ in favor of $s$, giving
a PWN field estimate of
$$
{\rm Log}(B_{10}) \approx 0.73 - 2/3 {\rm Log}(\tau_4) - 3.03 (s-3).
$$
The total energy of the electron population, $\approx 10^{47} 
B_{10}^{(s+1)/2} {\rm erg}$,
gives an upper limit on the initial pulsar period of 
$P_i \la 0.4 B_{10}^{-(s+1)/4}$s. Of course, adiabatic losses in the PWN flow
likely decrease the present electron energy and require smaller $P_i$.

\begin{inlinefigure}
\epsscale{0.8} \rotatebox{270}{\plotone{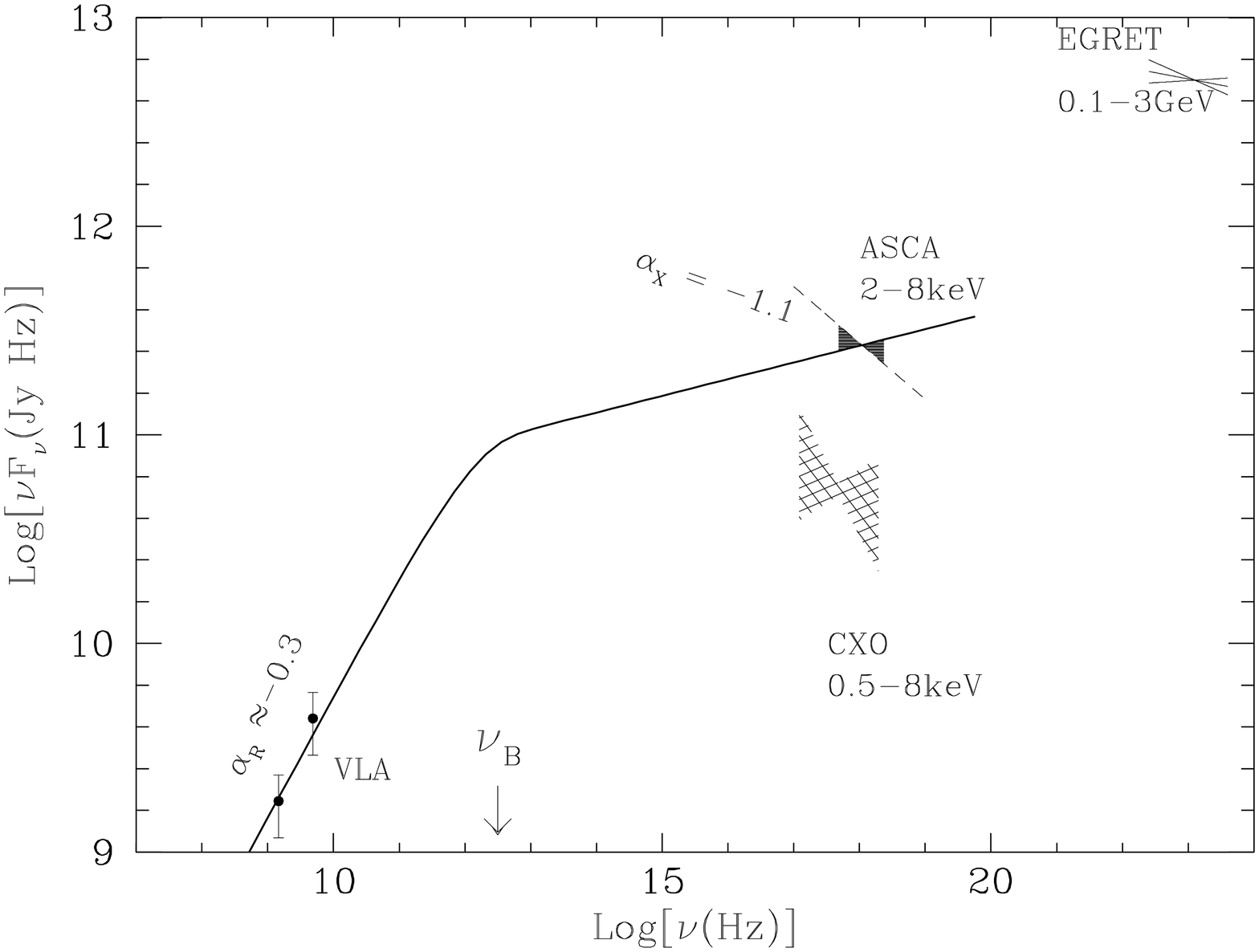}}
\figcaption{Our PWN candidate's broad-band SED, with X-ray error regions 
for the full nebula (\ASCA) and bright \CXO\ core.  The curve is the 
best-fit $s=2.85$ cooling synchrotron model.}
\label{SED}
\end{inlinefigure}

We can check consistency with another crude estimate of
the nebula field.  \citet{oka99} argued that the `fit' of the PWN candidate
with Lynds 227 suggests confinement and pressure equilibrium. The
CO line-width inferred pressure of $\sim 1.4 \times 10^{-11} {\rm g/cm/s^2}$
implies a nebular equipartition field of $\sim 20\mu$G. However, since
Lynds 227 brackets only on one side, the wind is likely only partly confined 
and $B$ may be larger.  For our best fit $s$, the equipartition $B$ 
suggests $\tau_4 \sim 20$, a Geminga-like pulsar. Our present poor
constraint on $\nu_B$ allows ages $\sim 3 \times 10^4 - 10^6$y,
including $B_{12}\tau_4 \approx 10$ as estimated above. If
the PWN $B$ exceeds $20\mu$G, even smaller $\tau$ are permitted.

	We can also follow the pulsar surface field $B_{12}$ out to the 
termination shock in a simple 1-D picture. The magnetic field at 
$r_s = \theta_s d$, the wind shock radius, is $B_s \sim 3 B_\ast 
r_\ast^3/(r_{LC}^2 r_s ) \sim 130 (B_{12} \tau_4 d_{1.8} 
\theta\arcmin )^{-1}\mu$G. Beyond $r_s$ the field evolution depends on
the poorly understood wind magnetization, but assuming $B_s \approx 20
\mu$G constant in the postshock wind, for $B_{12}\tau_4 \approx 10$ we 
expect termination at $\theta_s \approx 40\arcsec$. 
Our \ACIS\ image does not show a spherical wind with a subluminous zone 
at this $\theta_s$, but instead suggests a jet or pulsar trail, with 
the $\sim 15\arcsec$ knot in the jet/trail plausibly a termination shock.
The overall diffuse X-ray morphology suggests a pulsar trail
leading back to a PWN, as seen for PSR B1757-24 \citep{kasp01}, implying
a birthsite to the NW. However, the nebula might also be powered through
a jet, as for the asymmetric PSR 1509-58 PWN \citep[c.f.][]{gae01}.
In this case the faint extended emission to the SE (Figure 1) may be the
counterlobe. Under this interpretation, the S32 cluster at $r \approx 5 
d_{1.8}$pc provides a plausible pulsar birthsite, requiring a pulsar 
transverse velocity $v_\perp \approx 500 d_{1.8}/\tau_4$km/s.
Much higher S/N imaging or a proper motion measurement are needed to 
distinguish these possibilities.

The $\gamma$-ray flux is clearly an additional spectral component
(Figure 4), but its origin is puzzling. 3EG J1809-2328 (= GeV J1809-2327) 
is apparently one of the more variable Galactic plane sources \citep{tom99}.
\citet{oka99} suggested that the PWN-generated electrons
penetrate the molecular cloud, generating GeV photons via
relativistic Bremsstrahlung. As the PWN is well separated from the
molecular gas at $\sim (3\arcmin) d \sim 1.6$pc, one might expect variability
no faster than the $3l/c \sim 15$y flow-crossing time. This is 
substantially longer than the \EGRET\ variability timescale, suggesting an
origin closer to the pulsar. Inverse Compton emission from 
the PWN termination
shock seems attractive, but both Galactic IR and local synchrotron emission
fluxes fail by several orders of magnitude to account for the required
soft photon energy density. Even the optical/UV emission from the OB stars
in S32 contribute less than $10^{5.5}L_\odot$, failing to produce the
required target photon density at the $\sim 5$pc PWN distance by
$\ga 300 \times$. The $\gamma$-ray flux is quite plausible for 
magnetospheric (curvature) pulsar emission at the inferred 
${\dot E}$ and distance, but variability is unexpected from an 
isolated pulsar. Confirmation or exclusion of the $\gamma$-ray variability, 
together with detection or limits on $\gamma$-ray pulsations is the best
way to address this issue.
Unless an X-ray period can be measured, this will not be resolved until
until AGILE or GLAST make new sensitive $\gamma$-ray observations.

\acknowledgments

	This work was supported in part by CXO grant GO0-1125 (RWR) and a 
Quebec Merit Fellowship (MSER). We thank the referee for a detailed critique.

%\clearpage

\end{document}